\documentclass[sigconf,authorversion]{acmart}
\usepackage{enumitem}
\usepackage{tabularx}
\usepackage{subcaption}
\usepackage{multirow}
\usepackage{caption}
\usepackage[normalem]{ulem}
\usepackage{float}
\usepackage{colortbl} 
\usepackage[table]{xcolor} 
\usepackage{soul}

\setstcolor{red} 
\AtBeginDocument{%
  }

\setcopyright{acmlicensed}

\copyrightyear{2027}
\acmYear{2027}
\setcopyright{cc}
\setcctype{by}
\acmConference[SIGCSE TS 2027]{Proceedings of the 58th ACM Technical Symposium on Computer Science Education V.1}{February 17--20, 2027}{Sacramento, CA, USA}
\acmBooktitle{Proceedings of the 58th ACM Technical Symposium on Computer Science Education V.1 (SIGCSE TS 2027), February 17--20, 2027, Sacramento, CA, USA}
\acmDOI{10.1145/3857222.3860095}
\acmISBN{979-8-4007-3075-7/2027/02}

\newcommand{\gray}[1]{\textcolor{gray}{#1}}
\newcommand{\todo}[1]{#1}

\begin{document}


\title[Code, Commits, and Complexity in the GenAI Era]{Code, Commits, and Complexity in the GenAI Era: A~Longitudinal~Study~of~Team-Based~Capstone~Projects}

\author{Asma Shakil}
\orcid{0000-0003-4760-3792}
\affiliation{%
\institution{University of Auckland}
\city{Auckland}
\country{New Zealand}
}
\email{asma.shakil@auckland.ac.nz}

\author{Claus Brabrand}
\orcid{0000-0001-5093-5806}
\affiliation{
\institution{IT University of Copenhagen}
\city{Copenhagen}
\country{Denmark}}
\email{brabrand@itu.dk}

\author{Paul Denny}
\orcid{0000-0002-5150-9806}
\affiliation{
\institution{University of Auckland}
\city{Auckland}
\country{New Zealand}}
\email{paul@cs.auckland.ac.nz}

\renewcommand{\shortauthors}{Asma Shakil, Claus Brabrand, and Paul Denny}

\begin{abstract}

Generative AI tools are now widely available to students, but little is known about how team-based project codebases have changed since they appeared. This matters when project courses use repository activity as evidence of progress and learning. Existing studies have largely examined self-reports, controlled tasks, or short individual assignments, leaving open questions around how student codebases evolve in authentic project settings.  We present a ten-semester analysis of N=260 capstone team repositories from 2021 to 2025, spanning pre- and post-AI cohorts. We examine three dimensions: \emph{code size} (lines of code), \emph{code characteristics} (comment density, error-handling density, and cyclomatic complexity), and \emph{development behaviour} (commit volume, edit size, and commit frequency).
We find that end-of-semester codebases roughly triple, from about 3.4~KLOC in 2021--2023 to about 10~KLOC in 2024--2025. Code characteristics show a clear decoupling: cyclomatic complexity and error-handling density increase, while comment density does not. Later cohorts commit far more often and in greater volume, but individual commit sizes are largely unchanged, suggesting a shift in development rhythm rather than edit granularity. 
Most changes appear from 2024 onwards and are not uniformly monotonic. We did not directly observe AI use, so we report these as changes that coincide with increased GenAI availability rather than as effects of AI use. Our findings highlight the need to reconsider how student-produced artefacts are used and assessed in project courses.
\end{abstract}



\maketitle

\section{Introduction}
Generative AI (GenAI) tools, such as ChatGPT, Claude, and Gemini, are now widely available to students for programming tasks. These tools can generate, refactor, debug, and explain code on demand, raising key questions for education:
\begin{quote}
    \emph{How are student software development practices changing in the presence of GenAI, and how should educators interpret student-produced code in this new landscape?} 
\end{quote}
While early evidence suggests that GenAI tools may increase apparent productivity, they also complicate assumptions about effort, learning, and assessment~\cite{jovst2024impact, lyu2024evaluating}, and can expose misalignments between intended learning outcomes and assessment tasks \cite{brabrand2026constructive}.
Recent empirical work has begun to examine these changes. For example, Zhang et al.~\cite{zhang2026changes} report that students submit longer solutions, make larger edits, and iterate more frequently after the public release of ChatGPT, often without corresponding improvements in conceptual understanding or performance. However, such studies have mainly examined short-term, individual assignments with fixed specifications. Less is known about longer-running, open-ended, team-based projects, where students collaborate over many weeks, make architectural decisions, and build evolving codebases. \todo{Capstone projects therefore provide a useful context for examining how the size, structure, documentation, and commit history of student code have changed as GenAI tools have become available.}

\textbf{Pre-AI vs Post-AI.}
In this paper, we address these gaps through a longitudinal analysis of student capstone repositories across ten semesters (2021:S1--2025:S2), covering the \textbf{\emph{pre-AI} (2021--2022)} as well as the \textbf{\emph{post-AI} (2023--2025)} eras. 
\todo{This boundary follows the public release of ChatGPT on November 30, 2022~\cite{openai2022chatgpt}; since students adopted AI tools at different times, we treat it as approximate (Section~\ref{secn:context_study_design}).}
Using version-control histories from 260 team repositories, we analyse code size and code characteristics at two snapshots (mid- and end-semester), and development behaviour measured across the project window.

\textbf{Contributions.}
This paper makes four primary contributions:
\begin{enumerate}[leftmargin=*,topsep=0pt]
    \item \textbf{Longitudinal Analysis:} We provide one of the first multi-semester, repository-level analyses of capstone projects spanning the pre-AI and post-AI eras (260 repositories across ten semesters, from 2021--2025).
    \item \textbf{Measurement Framework:} We introduce a multi-dimensional measurement framework spanning i) \emph{code size}, ii) \emph{code characteristics}, and iii) \emph{development behaviour}, enabling a richer characterisation of AI-era change than single-metric studies permit.
    \item \textbf{Silent Complexity:} We identify a \emph{complexity-w/o-commentary} phenomenon in post-AI cohorts, where cyclomatic complexity and error-handling density rise while comment density remains flat---a finding with direct implications for code comprehension and assessment in AI-augmented codebases.
    \item \textbf{More, Not Bigger Commits:} \todo{Later cohorts commit far more often and in greater volume, yet commit size shows no clear trend. This shift in \emph{frequency, not granularity,} does not support the assumption that AI assistance yields fewer, larger ``batch'' commits.}
\end{enumerate}

\section {Related Work}
Our work sits at the intersection of three research strands: \emph{GenAI's impact on programming behaviour}, \emph{repository-level analysis of software development} in educational settings, and the \emph{assessment of code quality in student projects}. We review each in turn below.

\textbf{GenAI's Impact on Programming Behaviour.}
The rapid availability of GenAI tools has created new challenges for programming education~\cite{prather2023robots}. These tools can generate, explain, debug, and refactor code, making them highly relevant to tasks traditionally used to assess student understanding~\cite{bouvier2025rest}. Recent reviews highlight both benefits and risks: GenAI tools may support productivity and feedback, but may also encourage over-reliance and make it harder for educators to judge what students have learned~\cite{nathaniel2025literature, shen2026ai}.

Empirical studies have begun to examine these effects. Zhang et al.~\cite{zhang2026changes} provide a close point of comparison for our work, showing that after the release of ChatGPT, students submitted longer solutions, made larger edits between submissions, and iterated more frequently. However, their study focuses on fixed-assignment tasks, leaving open the question of whether similar changes occur in larger, open-ended projects. Other studies show that students use GenAI tools for both conceptual help and direct code generation~\cite{rahe2025programming}, and that tools such as GitHub Copilot can accelerate task progress while raising concerns about students' understanding of the generated suggestions~\cite{shihab2025effects}.

This body of work shows that GenAI tools can change student programming behaviour. However, most studies focus on short-term tasks, controlled settings, individual assignments, or self-reported use; less is known about how student codebases evolve over longer periods and across multiple cohorts. Our study addresses this gap through a five-year, ten-semester longitudinal repository-level analysis spanning pre-AI and post-AI cohorts.

\textbf{Repository Metrics in Education.}
Version-control data have long been used in computing education research to study how students work over time. Git and GitHub histories provide evidence of commit activity, contribution patterns, collaboration, and project progress. Prior studies have used these data to examine student development practices in software engineering and project-based courses. For example, Tushev et al.~\cite{tushev2020using} show that GitHub activity can provide useful insight into student work, while cautioning against treating simple activity metrics as direct measures of learning or performance. Hamer et al.~\cite{hamer2021using} use Git metrics to measure individual and team contributions, and Gitinabard et al.~\cite{gitinabard2020student} show that GitHub logs can reveal meaningful patterns in teamwork and development behaviour.

This prior work demonstrates the value of repository mining in educational contexts. However, much of it 
predates the widespread use of GenAI and focuses mainly on contribution, teamwork, or assessment. Our work builds on this tradition but applies it to longitudinal change across the pre- and post-GenAI eras. We combine behavioural metrics, such as commit volume, commit frequency, and edit size, with artefact-level metrics, such as lines of code, comment density, error-handling density, and cyclomatic complexity.

\textbf{Code Quality in Student Projects.}
A related stream of computing education research examines the quality and structure of student code. While correctness is often central to programming assessment, code quality is increasingly treated as an important learning outcome in its own right. Keuning et al.~\cite{keuning2023systematic} show that it is multi-dimensional, spanning readability, documentation, decomposition, complexity, maintainability, and style---and these are precisely the dimensions along which student code tends to fall short. In another study, Keuning et al.~\cite{keuning2017code} identify recurring quality problems in student programs and argue that students need support beyond simply producing working solutions, while Ihantola and Petersen~\cite{ihantola2019code} examine code complexity in introductory programming and highlight both the value and the limitations of applying traditional software metrics to student code. This same structural lens has more recently been turned on AI-generated code: in a controlled comparison of functionally equivalent programs, \citet{azeem2025ai} found that ChatGPT-generated code exhibited higher cyclomatic complexity and Halstead volume than human-written code while scoring lower on documentation and function structure.

These studies support the use of static metrics to characterise student-produced software, but they do not directly examine whether such metrics shift in the GenAI era. This matters because AI assistance may affect not only how much code students produce, but also how that code is structured, documented, and maintained. We therefore use multiple complementary metrics to understand how student codebases have changed over time: lines of code for scale, and cyclomatic complexity, error-handling density and comment density for code characteristics. These metrics describe properties of a codebase; they are not measures of code quality, and each has known limitations (Section~\ref{sec:code-characteristics}).

\section{Method}
\subsection{Context \& Study Design}
\label{secn:context_study_design}
Our analysis draws on GitHub repositories from our capstone course, a compulsory final-year course in our undergraduate computer science degree offered twice a year. Students working in teams of five to six build software for real external clients over a 13-week period, following an agile process and using GitHub for version control. The projects span a broad range of domains (web and mobile applications, AI systems, games, simulations, and desktop software) and technologies. \todo{JavaScript or TypeScript is the main language in 42--58\% of repositories in 2021--2023 and in 71--74\% in 2024--2025; Python leads in 2022 (26 of 65 repositories).} The working prototype is the primary assessment artefact.

Our data spans ten consecutive semesters (S1~2021--S2~2025). 
We divide the ten semesters into a pre-AI period (2021--2022) and a post-AI period (2023--2025), with the public release of ChatGPT in November 2022~\cite{openai2022chatgpt} as the boundary, after which general-purpose GenAI assistance became freely and widely available. The boundary is not sharp. GitHub Copilot was released as a technical preview in June 2021~\cite{github2021copilotPreview} and became free for verified students in June 2022~\cite{github2022copilotGA}, so some S2~2022 students may already have used AI code completion. 
Conversely, uptake after ChatGPT's release was gradual. For example, among UK undergraduates, the share using GenAI for assessed work rose from 53\% in late 2023 to 88\% in late 2024~\cite{freeman2025hepi}.
We therefore use ``pre-AI'' and ``post-AI'' as shorthand for periods of lower and higher GenAI availability, not as labels of actual use, and report per-cohort values so that the timing of change can be seen rather than assumed.
We define two fixed measurement (\emph{snapshot}) points within each semester:
\begin{itemize} 
    \item \textbf{Week-7} (\emph{mid}), following the mid-semester break, when teams present their first minimum viable product; \emph{and} 
    \item \textbf{Week-13} (\emph{end}), at the close of the project period, when teams present their final, complete solution. 
\end{itemize}
These milestones bracket the two main development phases: the \emph{mid}-point captures the project at its first functional state, and the \emph{end}-point captures the finished product. Comparing the two illustrates how projects evolve during their most active period. Snapshot dates are treated as end-of-day UTC cutoffs to ensure consistent and reproducible selection of commits across repositories.

\subsection{Research Questions}
Our study is guided by the following research questions:
\begin{itemize}[topsep=0pt,leftmargin=*]
    \item \textbf{RQ1 (Code Size):} How has the volume of code produced in capstone projects changed between the pre-AI era (2021–2022) and the post-AI era (2023–2025), as measured by lines of code?

    \item \textbf{RQ2 (Code Characteristics):} How have the qualitative characteristics of code produced in capstone projects changed between the pre-AI era and the post-AI era?

    \item \textbf{RQ3 (Coding Behaviour):} To what extent has the development behaviour of capstone students shifted between the pre-AI and the post-AI era?

\end{itemize}
RQ1--2 is about (code) \emph{product}; RQ3 is about (development) \emph{process}.

\subsection{Repository Selection \& Dataset}
\label{sec:dataset}
For each semester, we began with all repositories in the GitHub organisation for that cohort's capstone offering, yielding 302 repositories across ten semesters. We then excluded those not representing active student projects according to three criteria: \emph{empty or inactive} (no commits or no source code at any snapshot); \emph{non-project} (testing, demonstrations, or templates and identified through naming conventions and manual inspection); and \emph{duplicate} (forking or renaming, of which only one per team was retained). This produced a cleaned dataset of \textbf{N=260 team repositories}: 19 in 2021, 65 in 2022, 57 in 2023, 51 in 2024, and 68 in 2025.

For each week~7 or 13 snapshot, we use the latest commit on the main branch at or before the cutoff. Repositories with no commit before a cutoff are treated as missing and disregarded. This affects only Week-7, where six 2022 repositories had not yet committed, yielding N=254 at Week-7 (versus the full N=260 at Week-13). Metrics that require both snapshots, such as the mid-semester ratio (Section~\ref{sec:midsem-ratio}) and edit size (Section~\ref{sec:edit-size}), are computed only for repositories present in both snapshots. Further per-metric exclusions occur only when a metric is undefined, e.g.\ repositories with no functions for cyclomatic complexity.

\subsection{Code Size (RQ1)}
\label{sec:code-size}
To address RQ1, we measure both how much code teams produce and how early in the project that code accumulates.

\textbf{Mid \& End Code Size.}
\label{sec:loc}
We measure code size as the cumulative number of \emph{lines of code (LOC)} present in a repository at two fixed snapshots during the semester:
\begin{align*}
\textsc{MidCodeSize} &= \text{repository LOC at week 7 (\emph{mid})}; \\
\textsc{EndCodeSize} &= \text{repository LOC at week 13} (\emph{end}).
\end{align*}
These metrics capture the cumulative amount of source code present at the \emph{mid}-semester and \emph{end}-of-semester checkpoints, respectively.

At each resolved snapshot commit, the repository is checked out and analysed with \texttt{tokei}, a language-aware static analysis tool\footnote{\url{https://github.com/XAMPPRocky/tokei}}, which classifies each file's lines as \emph{code}, \emph{comment}, or \emph{blank} and ignores unrecognised file types.
\todo{We sum code lines across all detected languages into a cumulative LOC value per repository per snapshot, excluding dependency lock files (e.g., \texttt{package-lock.json}) and the dependency and build directories (e.g., \texttt{node\_modules}, virtual environments). Comment lines are retained separately for the comment-density metric (Section~\ref{sec:comment-density}).}
All values are tied to explicit snapshot commits, ensuring measurements are reproducible and independently verifiable.

\textbf{Mid-Semester Progress.}
\label{sec:midsem-ratio}
To capture not only \emph{how much} code teams produce but \emph{when} they produce it, we compute a mid-semester accumulation ratio as: 

\[
\textsc{MidSemesterProgress} = \frac{\textsc{MidCodeSize}}{\textsc{EndCodeSize}}.
\]
This metric isolates the \emph{timing} of code production independently of absolute codebase size.

\subsection{Code Characteristics (RQ2)}
\label{sec:code-characteristics}
\todo{We describe three size-normalised characteristics of each codebase at both snapshots (week 7 and week 13): control-flow complexity (mean per-function CCN), explicit error handling code (per KLOC), and the share of comment lines (relative to LOC). Prior work uses similar measures to characterise student and AI-generated code~\cite{azeem2025ai, keuning2023systematic}. We treat none of them as a direct measure of quality.}

\textbf{Cyclomatic Complexity.}
\label{sec:complexity}
To characterise the structural complexity of code, we compute function-level cyclomatic complexity (CCN). \todo{CCN reflects branching structure only and relates weakly to how hard code is to understand~\cite{feitelson2023}; we read it as a structural measure, not a measure of comprehensibility.} 
At each snapshot, we analyse source files in the codebase using \texttt{lizard} \footnote {Lizard (language-independent analysis tool): \url{https://github.com/terryyin/lizard}}, which supports several languages including Python, JavaScript/TypeScript, Java, and C\# and excludes vendor, build, and generated directories, minified bundles, and files exceeding 2~MB, under per-file and per-repository time limits. For each function \texttt{lizard} reports its CCN; the repository-level statistic is the \emph{mean} CCN across all functions, $f \in \mathbb{F}$ (the set of all functions):
\[
\textsc{MeanCyclomaticComplexity}
=
\frac{\sum_{f \in \mathbb{F}} \mathrm{CCN}(f)}
     {|\mathbb{F}|}.
\]

\textbf{Error-Handling Density.}
\label{sec:error-handling}
\todo{We measure how often code contains explicit error-handling constructs; more handling is not necessarily better, and we do not treat it as evidence of defensive design or of AI-generated code.}
For each resolved snapshot commit, the repository is checked out and its source files scanned for language-specific error-handling patterns.%
\footnote{Python \big(\texttt{try}/\texttt{except}/\texttt{raise}/\texttt{finally}\big); 
JavaScript \big(\texttt{try}/\texttt{catch}/\texttt{throw}/\texttt{.catch()}\big); 
Go \big(\texttt{if err != nil}/\texttt{return err}\big); 
Rust \big(\texttt{Result}/\texttt{match}/\texttt{unwrap}/\texttt{?}\big); etc.} Files under vendor or build directories, binary files, and files exceeding 2~MB are again excluded. A source line is counted once if it matches at least one pattern, and matched lines are summed across the repository. We normalise this count by the repository's LOC (Section ~\ref{sec:code-size}) to obtain error-handling density per thousand lines of code:
\[
\textsc{ErrorHandlingDensity} =
\frac{\text{\#Error-Handling Lines}}{\text{KLOC}}. 
\]
This metric is a lexical proxy: it captures the number of lines containing error-handling constructs rather than a semantic analysis of exception flow, and we treat it accordingly (Section~\ref{sec:threats}).

\textbf{Comment Density.}
\label{sec:comment-density}
\todo{Comment density captures the quantity of in-code commentary, not its quality, and more comments are not necessarily better, as comments may become inconsistent with the code~\cite{wen2019comments}.} We compute it as:
\[
\textsc{CommentDensity} =
\frac{\text{\#Comment Lines}}{\text{LOC}},
\]
where blank lines are excluded, \todo{and LOC is as defined in Section~\ref{sec:code-size}.}

\subsection{Coding Behaviour (RQ3)}
\label{sec:coding-behaviour}
To address RQ3, we characterise how students develop their code, including the size and cadence of commits that drive codebase evolution.

\textbf{Median Edit Size.}
\label{sec:edit-size}
To characterise how code changes are introduced, we compute \emph{edit size}, a commit-level metric that captures the magnitude of individual commits and, hence, the \emph{granularity} of development activity. For each repository, we consider all commits after the resolved Week-7 snapshot through and including the Week-12 snapshot commit on the default branch, a window that captures the latter half of the semester, when teams typically focus on feature completion, refactoring, and integration. We use the Week-12 boundary (rather than the Week-13 end-of-semester snapshot used elsewhere) so that edit size spans a fixed five-week development window; repositories with missing or identical Week-7 and Week-12 commits are excluded.

Edit size is computed per commit using Git's \texttt{diff -{}-numstat}, which reports lines added and deleted per file; binary files (for which line counts are undefined) are excluded. For a given commit~$C \in \mathbb{C}$ (in the set of all commits in the repository), edit size is the total number of modified lines across all affected files:
\begin{align*}
\textsc{EditSize}(C)
&=
\sum_{f \in C}
\left(\,
|\mathcal{A}(f)| + |\mathcal{D}(f)|
\,\right)
\\
\textsc{MedianEditSize}
&=
\operatorname{median}_{C \in \mathbb{C}}
\bigl(\,\textsc{EditSize}(C)\,\bigr),
\end{align*}
where $\mathcal{A}(f)$ and $\mathcal{D}(f)$ denote the sets of lines \emph{added to} and 
\emph{deleted from} file~$f$, respectively. \textsc{EditSize} therefore captures the total number of modified lines in a commit, whether the changes are additions, deletions, or refactorings.
For each repository, we summarise the per-commit edit sizes by their median, which is robust to the extreme outliers common in student repositories (e.g.\ bulk imports). We report the cohort median of these repository medians.

\textbf{Commit Volume.}
\label{sec:commits}
Alongside edit size, we characterise development cadence over the full project window on the default branch. \emph{Commit volume} is the total number of commits per repository:
\[
\textsc{CommitVolume} = |\mathbb{C}|.
\]

\textbf{Commit frequency} is the \emph{total number of commits per active day}, defined as the total commit count divided by the number of distinct (active) calendar days carrying at least one commit, $\mathbb{D}$:
\[
\textsc{CommitFrequency}
=
\frac{|\mathbb{C}|}
     {|\mathbb{D}|}.
\]
Commit frequency distinguishes teams that commit steadily from those that commit in concentrated sessions of the same total volume. We report the cohort median of each.

\subsection{Statistical Analysis}
\label{sec:stats}
We test each metric for a monotonic trend across the five cohorts (2021--2025), treating cohort year as an ordinal predictor at the repository level. We report Spearman's rank correlation ($\rho$) between cohort year and the per-repository value as the effect size, with the Theil--Sen slope as an interpretable estimate of change per year. To control the false discovery rate across the eight\footnote{\textsc{MidSemesterProgress} is deterministically derived from \textsc{MidCodeSize} and \textsc{EndCodeSize} and is therefore not counted as an additional (ninth) test.} metrics, we apply the Benjamini--Hochberg procedure, reporting both uncorrected ($p$) and adjusted ($q$) values. As repositories are nested within semesters and
teams, we restrict inference to repository-level trends and do not model within-team dependence.

\section{Results}\label{sec:results}
We examine whether each metric changes consistently across the five cohorts (2021---2025). For each metric, we treat cohort year as an ordered variable and test for a monotonic trend at the repository level. We report Spearman's rank correlation ($\rho$) as the effect size, and use the Benjamini--Hochberg procedure to control the false discovery rate across all metrics ($q$). Snapshot metrics are measured at the end-of-semester checkpoint, week~13 ($N=260$ repositories). 
Behavioural metrics are measured across the full project period. Full statistics are shown in Table~\ref{tab:trends}.

\begin{table*}[t]\centering\small
\caption{Per-cohort medians and five-cohort monotonic-trend statistics, grouped by research question. $\star$ marks trends significant after Benjamini--Hochberg correction ($q \leq 0.05$); non-significant trends are in grey.}
\label{tab:trends}
\begin{tabular}{l l c rrrrr r r r r}
\toprule
 & & & \multicolumn{5}{c}{Median by cohort} & & & & \\ \cmidrule(lr){4-8}
\textbf{RQ} & \textbf{Metric} & & 2021 & 2022 & 2023 & 2024 & 2025 & $\rho$ & Slope/yr & $p$ & $q$ \\
\midrule
\multirow{3}{*}{\textbf{RQ1 (Code Size)}}
 & \textsc{MidCodeSize (KLOC)}            & $\star$ & 1.5 & 1.4 & 1.8 & 2.5 & 4.8 & $\mathbf{+0.34}$ & $+0.66$ & $<$0.001 & $<$0.001 \\
 & \textsc{EndCodeSize (KLOC)}            & $\star$ & 2.3 & 3.7 & 3.4 & 7.7 & 11 & $\mathbf{+0.45}$ & $+2.0$ & $<$0.001 & $<$0.001 \\
 & \gray{\textsc{MidSemesterProgress} (\%)} & & \gray{73} & \gray{46} & \gray{58} & \gray{40} & \gray{51} & \gray{$-0.12$} & \gray{$-2.8$} & \gray{0.069} & \gray{0.092} \\
\midrule
\multirow{3}{*}{\textbf{RQ2 (Code Characteristics)}}
 & \textsc{MeanCyclomaticComplexity} & $\star$         & 1.7 & 2.0 & 1.8 & 2.2 & 2.5 & $\mathbf{+0.27}$ & $+0.17$ & $<$0.001 & $<$0.001 \\
 & \textsc{ErrorHandlingDensity} (/KLOC) & $\star$ & 10 & 3.7 & 4.8 & 12 & 13 & $\mathbf{+0.27}$ & $+1.9$ & $<$0.001 & $<$0.001 \\
 & \gray{\textsc{CommentDensity} (\%)}         & & \gray{9.2} & \gray{8.6} & \gray{7.6} & \gray{7.6} & \gray{8.4} & \gray{$-0.04$} & \gray{$-0.19$} & \gray{0.52} & \gray{0.52} \\
\midrule
\multirow{3}{*}{\textbf{RQ3 (Coding Behaviour)}}
 & \gray{\textsc{MedianEditSize}} &           & \gray{65} & \gray{38} & \gray{61} & \gray{51} & \gray{65} & \gray{$+0.09$} & \gray{$+3.0$} & \gray{0.14} & \gray{0.16} \\
 & \textsc{CommitVolume} & $\star$      & 63 & 112 & 100 & 265 & 300 & $\mathbf{+0.38}$ & $+46$ & $<$0.001 & $<$0.001 \\
 & \textsc{CommitFrequency} & $\star$ & 3.3 & 4.2 & 4.8 & 5.3 & 5.4 & $\mathbf{+0.26}$ & $+0.45$ & $<$0.001 & $<$0.001 \\
\bottomrule
\end{tabular}
\end{table*}

\subsection{Code Size (RQ1)}

\begin{figure}[t]
 \centering
 \includegraphics[width=0.9\linewidth]{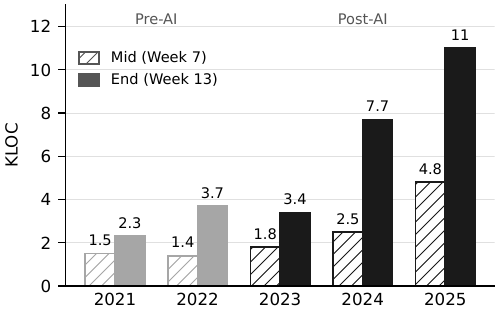}
 \caption{Median KLOC per repository (RQ1) at mid (Week~7) and end (Week~13)  snapshots (\textcolor{gray} {$\blacksquare$} pre-AI; \textcolor{black}{$\blacksquare$} post-AI).}
 \label{fig:loc}
\end{figure}

End-of-semester code size increases substantially across the study period (Figure \ref{fig:loc}, top rows of Table~\ref{tab:trends}). 
\todo{The median is 2.3~KLOC in 2021, 3.7 in 2022 and 3.4 in 2023, then rises to 7.7~KLOC in 2024 and 11~KLOC in 2025; the 2024--2025 median is roughly three times the 2021--2023 median. The trend is statistically significant ($\rho = +0.45$, $q < 0.001$), with a Theil--Sen slope of about +2.0~KLOC per year. Mid-semester code size follows the same pattern, rising from 1.4--1.8~KLOC in 2021--2023 to 2.5~KLOC in 2024 and 4.8~KLOC in 2025.}

\todo{The proportion of final code present by mid-semester (week~7) shows no clear trend ($\rho = -0.12$, $p = 0.069$): later cohorts produce more code overall, but not proportionately earlier in the semester.}

\begin{quote}
\textbf{\textsc{\protect\underline{Result 1}} (Code volume roughly triples from 2024)}: Median end-of-semester repository size rose from \textbf{3.4~KLOC in 2021--2023 to 10~KLOC in 2024--2025}, with no increase in 2023. 
\end{quote}

\subsection{Code Characteristics (RQ2)}

The three code-characteristic metrics show different patterns (middle rows of Table~\ref{tab:trends}). Cyclomatic complexity increases steadily across the cohorts. Mean per-function CCN rises from 1.7 in 2021 to 2.5 in 2025 ($\rho=+0.27$, $q<0.001$). This suggests that student code becomes more structurally complex over time.

Error-handling density also increases, although the pattern is less smooth. It falls from 10 per KLOC in 2021 to 3.7 in 2022 and 4.8 in 2023, then rises to about 12--13 in 2024--2025. This trend is also statistically significant ($\rho = +0.27$, $q < 0.001$).

\todo{Comment density, in contrast, shows no clear trend ($\rho = -0.04$, $p = 0.52$). It is highest in 2021 (9.2\%), lowest in 2023--2024 (7.6\%) and 8.4\% in 2025, an uneven pattern rather than a steady decline.}

Overall, student code becomes larger, more complex and contains more error-handling constructs, but comment density does not increase.

\begin{quote}
\textbf{\textsc{\underline{Result 2}} (Complexity without commentary):}
Cyclomatic complexity and error-handling density both increase significantly. Comment density shows \textbf{no clear trend}. Post-AI code is \textbf{more complex and contains more error-handling constructs, but is not more documented}.
\end{quote}

\subsection{Coding Behaviour (RQ3)}

Coding activity also increases across the study period (bottom rows of Table~\ref{tab:trends}). The median number of commits per repository rises from 63 in 2021 and around 100--112 in the 2022--2023 cohorts to 265 in 2024 and 300 in 2025. Commit volume shows the strongest trend among the behavioural metrics ($\rho=+0.38$, $q<0.001$).

Commit frequency also increases. The median number of commits per active day rises from 3.3 in 2021 to 5.4 in 2025 ($\rho=+0.26$, $q<0.001$).
However, the size of each commit does not change much. The median number of lines changed per commit after mid-semester shows no clear trend across the 2021--2025 cohorts ($\rho=+0.09$, $p=0.14$; cohort medians 38--65 lines). This result holds even when repositories with very little activity are excluded.
This indicates that overall later cohorts commit much more often and produce much more code, but they do so through a larger number of similarly sized commits, rather than through larger commits.

\begin{quote}
\textbf{\textsc{\underline{Result 3}} (More commits, not bigger ones):}
Median commit volume \textbf{tripled}, increasing from around 100 commits to around 300 commits per repository. Commits per active day also rose from \textbf{3.3 to 5.4}. Post-AI students commit \textbf{more often, but not in larger batches}.
\end{quote}

\section{Discussion}
\label{sec:discussion}

Our results suggest that the GenAI-era change in capstone development is best understood as multi-dimensional: codebases have grown, development activity has intensified, and code structure has shifted, but these changes occur unevenly across metrics.

\textbf{The Shape of Change: Large, \todo{Later}, and Uneven.}
By the end of the semester, \todo{repositories in 2024--2025 are about three times the size of those in 2021--2023}. This extends the pattern \citet{zhang2026changes} report from short individual assignments to longer-running team projects, and is consistent with evidence that AI assistance can increase coding throughput \citep{peng2023impact}. However, the change is not abrupt: code size, complexity, error-handling density, and commit volume change little in 2023 and then rise from 2024 onward. 
The 2023 cohort appears transitional: it was the first after the public release of ChatGPT, but use of AI tools was likely still uneven (Section~\ref{secn:context_study_design}).

\textbf{Complexity Without Commentary.}
The code-characteristic metrics split: cyclomatic complexity and error-handling density rise, but comment density does not. Later cohorts thus submit code that is larger, more complex, and contains more error-handling constructs, but not better documented. This is consistent with evidence that AI-assisted code can increase structural complexity without a commensurate gain in maintainability: \citet{azeem2025ai} found ChatGPT-generated code to be more complex and less well-documented than human-written code. The concern is not that students produce more code, but that it may be harder to inspect and maintain---sharpened by evidence that heavy reliance on AI can impair the very skills needed to review it \citep{shen2026ai}. In project courses, documentation, traceability, and the ability to explain one's own code may therefore need to become explicit assessment criteria, through viva-style defences, code walkthroughs, or stronger documentation requirements.

\textbf{A Behavioural Signature of Frequency, Not Granularity.}
Later cohorts commit far more often, both in total and per active day, yet median commit size does not increase. This challenges the expectation that AI-generated code would appear as fewer, larger commits: instead, teams produce more code through more frequent, similarly sized commits. This differs from the assignment-level pattern in \citet{zhang2026changes}, suggesting development behaviour may vary between short individual tasks and longer team projects, and aligns with \citet{shihab2025effects}, who found Copilot shaped students' processes, not just their outputs. AI assistance may speed how quickly students integrate and revise code without changing commit size, though as our study is correlational, we treat this as a hypothesis rather than a causal claim.

\textbf{Implications for Assessing AI-Era Artefacts.}
Our findings complicate artefact-based assessment. Metrics that educators may once have treated as signs of effort or progress, such as code volume and commit activity, have increased substantially, while comment density has not. A larger, more complex, more frequently committed codebase is therefore not necessarily evidence of greater learning. This reinforces concerns that traditional programming assessments become less reliable under GenAI: \citet{prather2023robots} argue for rethinking assessment, and \citet{bouvier2025rest} emphasise assessing students' reasoning, process, and ability to explain their work. This does not mean AI use is harmful; rather, assessment is likely more robust when the final codebase is weighed alongside process evidence and students' ability to justify their work.

\section{Threats to Validity}
\label{sec:threats}
This study has several limitations. 
First, we do not directly observe students' use of GenAI. As discussed in Section~\ref{secn:context_study_design}, the pre/post boundary marks a change in availability rather than use, so our findings are observed changes during a period of increasing AI availability, not causal evidence of AI use.
Other changes over the same period include changes in course staff, assessment criteria, and project types across semesters. The 2021 and some of the 2022 cohorts worked partly remotely under COVID-19 restrictions, and the 2021 cohort is much smaller (19 repositories) than later ones. Changes in frameworks and project templates could also add boilerplate code independently of AI use. Our design cannot separate these factors from changes in GenAI availability.

\todo{Second, as noted in Section~\ref{sec:code-characteristics}, our code-characteristic metrics are descriptive proxies.} For example, our error-handling metric is lexical rather than semantic, and may count language-specific idioms, comments, or strings that do not represent \todo{genuine error handling}. The language mix shifted towards JavaScript/TypeScript over the study period (see Section~\ref{secn:context_study_design}), but restricting the analysis to the 157 repositories whose main language is JavaScript or TypeScript, all five significant trends hold (e.g., \textsc{EndCodeSize} $\rho = +0.42$, $q < 0.001$).
Also, LOC and comment density are derived from \texttt{tokei}, which covers a broader range of languages, whereas cyclomatic complexity is derived from \texttt{lizard}, which supports only a subset. The \emph{complexity-without-commentary} pattern, which pairs a \texttt{lizard}-based measure with a \texttt{tokei}-based one, should therefore be interpreted at the cohort level, rather than as a direct within-file comparison.

Finally, the findings come from a single team-based capstone course at a single institution. They may not transfer to courses with different assessment structures, languages, or student populations. Repositories are also nested within semesters and teams, and our analysis treats repository-level trends descriptively rather than modelling all sources of dependence. Although we control the false discovery rate across metrics, the monotonic trend tests have limited sensitivity to non-monotonic patterns, \todo{such as a low in 2022--2023 followed by a rise from 2024}. Null results should therefore be read as the absence of a clear monotonic trend, not as evidence that no change occurred.

\section{Conclusion}
\label{sec:conclusion}
We have presented a ten-semester, repository-level analysis of 260 student capstone projects from 2021 to 2025, examining changes in student code and development practices during the rise of GenAI. We find substantial shifts across code size, code characteristics, and development behaviour. \todo{End-of-semester codebases in 2024--2025 are roughly three times the size of those in 2021--2023;} cyclomatic complexity and error-handling density increase, while comment density does not. Students also commit more frequently, though individual commit sizes remain largely unchanged.

These findings suggest that capstone project work \todo{since 2023} is characterised by larger, more complex codebases that contain more \todo{error-handling constructs}, but are not necessarily better documented. This \emph{complexity-without-commentary} pattern may make student code harder to understand and assess. Similarly, increased commit activity reflects \emph{frequency rather than granularity}, meaning traditional indicators such as code volume and commit counts may no longer map cleanly onto student effort or learning. \todo{As AI use was not observed, these are changes that coincide with increased GenAI availability, not demonstrated effects of it.}
Future work should link repository-level trends to observed AI use, distinguish AI-authored from human-authored code, and examine whether rising complexity reflects sound engineering practice or the verbose style often associated with GenAI output.

\section*{\todo{Acknowledgments}}
\todo{We are grateful to reviewers for their many helpful comments, especially regarding our metrics, that have enabled us to strengthen this work.}

\bibliographystyle{ACM-Reference-Format}
\balance 
\bibliography{sample-base}

\end{document}